\documentstyle[12pt]{article}
\begin{document}

\begin{center}

{\Large \bf A local approach for global partial density of states}

\bigskip

Jian Wang and Qingrong Zheng

\bigskip

{\it
Department of Physics, \\
The University of Hong Kong,\\
Pokfulam Road, Hong Kong.
}

\bigskip

Hong Guo

\bigskip

{\it
Centre for the Physics of Materials,\\
Department of Physics, McGill University,\\
Montreal, Quebec, Canada H3A 2T8.
}

\end{center}

\vfill

\baselineskip 15pt               

To apply the scattering approach for the problem of AC transport 
through coherent quantum conductors, various partial density of states
must be evaluated. If the global partial density of states (GPDOS)
is calculated externally using the energy derivatives of the scattering
matrix, the results are not precise unless the conductor has a 
large scattering volume. We propose a local formula for GPDOS which is 
suitable for any finite scattering volume.  We apply this formula to 
compute the emittance of a two-dimensional quantum wire under the 
multi-mode and finite temperature condition.

\vfill

\baselineskip 16pt


\newpage

The dynamic conductance of a quantum coherent mesoscopic system under 
a time dependent external field is the subject of recent 
interests\cite{but1,bruder,pieper,chen,wang1,wang2}. In contrast to 
dc-transport in the linear regime, where the internal potential distribution 
inside a sample does not appear explicitly, the AC-response depends 
sensitively on the internal potential distribution. This internal potential 
is due to the charge distribution generated by the applied external AC-field 
at the leads and it has to be determined self-consistently\cite{but1}. 
A particular useful approach to investigate AC transport properties of
coherent quantum conductors, is to study the AC-response of the system
to an external perturbation which prescribes the potentials in the reservoirs 
only\cite{pastawski,but1}. The external potentials effectively determine the 
chemical potential of the reservoirs and the potential distribution in the 
conductor must be considered a part of the response which is to be calculated
self-consistently. In this approach, B\"uttiker and his 
coworkers\cite{but1,but3} have formulated a current conserving 
formalism for the low frequency admittance of mesoscopic conductors. 

In the theory of B\"uttiker, Pr\^etre and Thomas\cite{but1}, it is 
necessary to consider the Coulomb interactions between the many 
charges inside the sample, in order to preserve the current conservation. 
For a multi-probe conductor the low frequency admittance is found to 
have the form\cite{but3,but4}
$G_{\alpha\beta}(\omega)=G_{\alpha\beta}(0)-i\omega E_{\alpha\beta}
+O(\omega^2)$, 
where $G_{\alpha \beta}(0)$ is the dc-conductance, $E_{\alpha \beta}$
is the emittance\cite{but3}, and $\alpha$ (or $\beta$) labels the probe. 
The emittance $E_{\alpha \beta}$ describes the current response at probe 
$\alpha$ due to a variation of the electro-chemical potential at probe 
$\beta$ to leading order with respect to frequency $\omega$. It can 
be written as\cite{but3}
\begin{equation}
E_{\alpha \beta} = \frac{dN_{\alpha \beta}}{dE} - D_{\alpha \beta},
\label{eq0}
\end{equation}
where the term $dN_{\alpha \beta}/dE$ is the global partial density of 
states (GPDOS)\cite{gas1} which is related to the scattering matrix. 
It describes the density of states of carriers injected in probe $\beta$ 
reaching probe $\alpha$ and is due to the response to the external 
perturbation. The term $D_{\alpha \beta}$ is due to the Coulomb interaction 
of electrons inside the sample and is the response to the internal
potential. $D_{\alpha\beta}$ 
can be computed from the local density of states\cite{but1,but3} 
which is related to the electron dwell times through the relation
$\sum_{\alpha} D_{\alpha \beta} = \tau_{d \beta}/h$
where $\tau_{d \beta}$ is the dwell time for particles coming from the 
probe $\beta$. Electric current conservation, namely
$\sum_{\alpha}G_{\alpha\beta}(\omega)=0$, means that 
$\sum_{\alpha}E_{\alpha\beta}=0$ or equivalently\cite{but1,ian}
\begin{equation}
\frac{dN_{\beta}}{dE}\equiv \sum_{\alpha} \frac{dN_{\alpha \beta}}{dE} 
= \sum_{\alpha} D_{\alpha\beta}
\label{conserve}
\end{equation}
where $dN_{\beta}/dE$ is the DOS for electron coming from the probe
$\beta$. Clearly the current conservation is 
established since one realizes that $\sum_{\alpha}dN_{\alpha\beta}/dE$ is 
the physical quantity called injectance which is identical\cite{but3} to 
$\sum_{\alpha}D_{\alpha\beta}$.

The physical meaning and the important role played by the various partial 
density of states (PDOS) is the subjects of extensive discussions\cite{gas1}.
While in one-dimensional (1D) systems the PDOS can be evaluated
analytically via the help of scattering Green's function, in 2D one is
usually forced to use numerical methods due to the complexity of the
problem\cite{wang1}, except for very special and exactly solvable
cases\cite{wang3}. In the AC transport formalism outlined above, the 
GPDOS can be expressed {\it approximately} in terms of the energy 
derivative of the scattering matrix elements\cite{avishai}: 
\begin{equation}
\frac{dN_{\alpha\beta}}{dE} = \frac{1}{4\pi i} \int dE (-\frac{df}{dE})
Tr \left[ {\bf s}_{\alpha\beta}^{\dagger}\frac{d{\bf s}_{\alpha \beta}}{dE} - 
\frac{d{\bf s}_{\alpha\beta}^{\dagger}}{dE}{\bf s}_{\alpha\beta}\right]\ \ .
\label{eq2}
\end{equation}
In Eq.(\ref{eq2}) $f(E)$ is Fermi distribution function and ${\bf s}_{\alpha 
\beta}$ is the scattering matrix of dimensions $M_{\alpha} \times M_{\beta}$
where $M_{\alpha}$ is the number of quantum channels supported by the 
lead $\alpha$. Because for a given system one may be able to obtain the 
scattering matrix, Eq.(\ref{eq2}) thus provides a practical means of 
computing the GPDOS. 

Using explicit 2D examples, it has been numerically demonstrated in 
Ref. \cite{zheng} and analytically shown in Ref. \cite{wang2}, that
the DOS $dN_{\beta}/dE$ computed from Eq.(\ref{eq2}) is only accurate
up to correction terms which are exponentially decaying functions of the
scattering volume away from thresholds of successive transport channels. At
the thresholds, the correction terms diverge. Hence precise current 
conservation can not be obtained from the explicit 
calculations\cite{wang1,wang2,zheng} 
until the correction terms are added. A more serious problem is 
that we only know how to correct the DOS since we can compare the left and 
right hand sides of Eq. (\ref{conserve}), but we don't know how to distribute
the correction terms among the PDOS of Eq. (\ref{eq2}). 
The whole issue comes about since the external response 
in the theoretical formalism is really formulated 
with a time dependent perturbation at minus or plus infinity which are 
asymptotically far from the scattering region, while for every practical 
calculation one wants to consider what happens in some finite 
scattering volume\cite{but5}.

The purpose of the present work is to formulate a procedure which allows 
a precise determination of the GPDOS $dN_{\alpha\beta}/dE$ for any finite 
scattering volume. We have used a general approach similar to that employed 
by Christen and B\"uttiker\cite{but6} in computing the non-linear transport 
coefficients by eliminating the GPDOS. 
This is possible if we assume that electric current conservation 
is satisfied. In doing so all the required quantities for the emittance 
become local, thus can be calculated precisely using the new formula 
(see below) within a finite scattering volume. We have applied this formula
to compute the emittance of a T-shaped quantum wire under the multi-mode
and finite temperature condition.

To proceed we recall from our previous investigations\cite{wang2,zheng} of
2D quantum conductors, that the discrepancy between the DOS as computed 
externally or internally is related to the mode mixing of the 2D scattering.
In 2D situations complicated mode mixing takes place.  This mode mixing
generates evanescent modes which can not propagate in the leads. For a 
scattering volume with a finite linear size $L$, the evanescent modes may
``leak'' out of the volume.  However when we calculate the GPDOS from
the scattering matrix using Eq. (\ref{eq2}), these ``leaked'' evanescent 
modes are not explicitly included, leading to a slightly inaccurate
calculation. On the other hand,  when we compute the local density of states
{\it internally} (see below) using the scattering wavefunctions, all the
modes, including the evanescent modes, are included. 
Indeed, as emphasized by B\"uttiker\cite{but3}, the AC transport formalism
guarantees electric current conservation when the scattering volume is
large enough to ensure that there is no electric field lines penetrating 
the surface of the volume.  This condition is certainly violated
due to the ``leaked'' evanescent modes when the volume is small.
Since evanescent modes do not contribute to electric current (but does 
to the DOS), it seems to be natural to use the conservation law to 
eliminate the need of computing GPDOS {\it externally}. This will be our
approach.

Using the electric current conservation relation of Eq.
(\ref{conserve}) and the injectance formula\cite{but3} of
$\sum_{\alpha}D_{\alpha\beta}$, we have
\begin{equation}
\sum_{\alpha}\left[\frac{dN_{\alpha\beta}}{dE}-D_{\alpha\beta}\right]
\ =\ \sum_{\alpha}\left[\frac{dN_{\alpha\beta}}{dE}-\int
\frac{dn_{\alpha\beta}}{dE}d^3{\bf r}\right]\ =\ 0\ .
\label{conserve1}
\end{equation} 
This is consistent with the relationship between the GPDOS and local 
PDOS (LPDOS)
\begin{equation}
\frac{dN_{\alpha \beta}}{dE} = \int \frac{dn_{\alpha \beta}({\bf r})}{dE} 
d^3{\bf r}
\label{e1}
\end{equation}
where 
\begin{equation}
\frac{dn_{\alpha \beta}({\bf r})}{dE} = -\frac{1}{4\pi i} \int dE
(-\frac{df}{dE}) Tr \left[
{\bf s}_{\alpha \beta}^{\dagger} 
\frac{\delta {\bf s}_{\alpha \beta}}{e \delta U({\bf r})}  - 
\frac{\delta {\bf s}_{\alpha \beta}^{\dagger}}{e \delta U({\bf r})} 
{\bf s}_{\alpha \beta} \right] \ \ .
\label{lpdos}
\end{equation}
is the LPDOS\cite{gas1}. Hence the GPDOS can be computed {\it locally}
through Eq.(\ref{lpdos}) if we can obtain LPDOS.
In general, for 2D systems it is very difficult if not impossible to obtain 
LPDOS using Eq. (\ref{lpdos}) since a complicated functional derivative must 
be evaluated. In 2D this functional derivative can only be computed for 
specially simple systems\cite{wang3}.  

For 1D systems, on the other hand, as shown in Ref. \cite{gas1}
simplification to the LPDOS formula Eq. (\ref{lpdos}) can be obtained   
(see Ref.\cite{gas1} for details) via the Fisher-Lee relation\cite{lee} 
between the scattering matrix and the retarded Green's function:
\begin{equation}
s_{\alpha \beta} = -\delta_{\alpha \beta} +i\hbar \sqrt{v_{\alpha} v_{\beta}}
G(x_{\alpha},x_{\beta})
\label{leefisher}
\end{equation}
where $x_{\alpha}$ is the boundary of the scattering region. 
The functional derivative of the Green's function $\delta G/\delta U$ is
given by\cite{gas1}
\begin{equation}
\frac{\delta G(x_{\alpha},x_{\beta})}{\delta U(x)} = G(x_{\alpha},x)
G(x,x_{\beta})\ \ \ .
\label{dgdu}
\end{equation}
Furthermore one can prove that for 1D systems the following relation is
true\cite{gas1},
\begin{equation}
G(x_1,x) G(x,x_2) = G(x_1,x_2) G(x,x)
\label{relation}
\end{equation}
for $x_1<x<x_2$.  Using Eqs. (\ref{leefisher}) - (\ref{relation}), it is not
difficult to derive\cite{gas1}, for 1D systems,  the following expressions
for the LPDOS
\begin{equation}
\frac{dn_{\alpha \beta}({\bf r})}{dE} = \frac{1}{2} T_{\alpha \beta} 
\frac{dn({\bf r})}{dE}
\label{e4}
\end{equation}
for $\alpha \neq \beta$ and
\begin{equation}
\frac{dn_{\beta \beta}({\bf r})}{dE} = \frac{d\bar{n}_{\beta}({\bf r})}{dE} 
-\frac{1}{2} \sum_{\alpha \neq \beta} T_{\alpha \beta} \frac{dn({\bf r})}{dE} 
\label{e5}
\end{equation}
where $T_{\alpha \beta}$ is the transmission coefficient from lead
$\beta$ to $\alpha$. In these results, the local DOS 
$d\bar{n}_{\beta}({\bf r})/dE$ is defined as
\begin{equation}
\frac{d\bar{n}_{\beta}({\bf r})}{dE} = \sum_{\alpha} 
\frac{dn_{\alpha \beta}({\bf r})}{dE}
\end{equation}
which is called the injectivity and it measures the additional local charge 
density brought into the sample at point ${\bf r}$ by the oscillating chemical 
potential at probe $\beta$. In general, the injectivity can be expressed 
in terms of the scattering wavefunction as\cite{but1}
\begin{equation}
\frac{d\bar{n}_{\beta}({\bf r})}{dE} = \int dE (-\frac{df}{dE}) 
\sum_n \frac{|\Psi_{\beta n}({\bf r})|^2} {h v_{\beta n}} \ \ ,
\label{dn}
\end{equation}
where $v_{\beta n}$ is the velocity of carriers at the Fermi energy at mode 
$n$ in probe $\beta$.  A related quantity, 
$d\underline{n}_{\alpha}({\bf r})/dE$, called emissivity, 
describes the local density of states of carriers at point ${\bf r}$ which 
are emitted by the conductor at probe $\alpha$. It is defined as
\begin{equation}
\frac{d\underline{n}_{\alpha}({\bf r})}{dE} = \sum_{\beta} 
\frac{dn_{\alpha \beta}({\bf r})}{dE} \ \ .
\label{emissivity}
\end{equation}
It has been shown\cite{but4} that in the absence of a magnetic field the
injectivity is equal to the emissivity. In the presence of a magnetic 
field the microreversibility of the scattering matrix implies that the 
emissivity into contact $\alpha$ in magnetic field $B$ is equal to the 
injectivity of contact $\alpha$ if the magnetic field is reversed\cite{but4},

\begin{equation}
\frac{d\underline{n}_{\alpha}({\bf r}, B)}{dE} = \frac{d\bar{n}_{\alpha}
({\bf r}, -B)}{dE} .
\label{micro}
\end{equation}
Finally, $dn({\bf r})/dE$ is the 
local DOS given by
\begin{equation}
\frac{dn({\bf r})}{dE} = \sum_{\alpha \beta} \frac{dn_{\alpha \beta}
({\bf r})}{dE}\ \ .
\label{ldos}
\end{equation}
From these results, one is able to calculate the LPDOS via Eq. (\ref{e4})
and (\ref{e5}) using the scattering wavefunctions.

For 2D systems, the Fisher-Lee relation is of the 
form\cite{datta}
\begin{equation}
s_{\alpha n \beta m} = -\delta_{\alpha n \beta m} +i\hbar \sqrt{v_{\alpha n} 
v_{\beta m}} \int \int G(x_{\alpha},y_{\alpha},x_{\beta},y_{\beta})
\chi_{\alpha n}(y_{\alpha}) \chi_{\beta m}(y_{\beta}) dy_{\alpha} dy_{\beta} 
\end{equation}
where $\chi_{\alpha n}$ is the transverse wavefunction in lead $\alpha$. 
In 2D, the equations similar to Eq. (\ref{e4}) and (\ref{e5}) do not seem 
to apply. This is because Eq.(\ref{relation}) does not hold in 2D\cite{foot2}.
However, since in a general 2D case one can not obtain analytical
expressions for the scattering matrix and numerical computation are usually
needed, it is thus enough to have a numerical prescription for calculating
the LPDOS. Our numerical method makes use of a mathematical identity
\begin{equation}
\frac{d{\bf s}_{\alpha \beta}}{dV} = \int d^3 r \frac{\delta {\bf s}_
{\alpha \beta}} {\delta U({\bf r})}\ \ \ ,
\label{e6}
\end{equation}
where $d{\bf s}_{\alpha \beta}/dV$ is calculated as 
follows\cite{leavens,but4,wang3}:
adding a constant potential $V$ in the scattering volume and computing the
scattering matrix formally to get ${\bf s}_{\alpha\beta} = {\bf s}_{\alpha 
\beta}(V)$, then taking the derivative and putting $V=0$. Numerically the 
derivative can be easily carried out using finite differencing.

Using Eqs.(\ref{e6}), (\ref{e1}), and (\ref{lpdos}), the GPDOS is completely
expressed by local quantities determined inside the scattering volume thus
can be computed accurately for any system sizes,
\begin{equation}
\frac{dN_{\alpha\beta}}{dE} = -\frac{1}{4\pi i} \int dE (-\frac{df}{dE})
Tr \left[ {\bf s}^{\dagger}_{\alpha \beta}
\frac{d{\bf s}_{\alpha \beta}}{dV} - \frac{d{\bf s}^{\dagger}_{\alpha \beta}}
{dV} {\bf s}_{\alpha \beta} \right]\ \ .
\label{gpdos}
\end{equation}
This equation is also valid in 1D. In the following, we shall compute the
emittance from Eq. (\ref{eq0}) using the Eq.(\ref{gpdos}) for GPDOS. To 
obtain the quantity $D_{\alpha\beta}$ we shall use the Thomas-Fermi 
approximation, in which case $D_{\alpha\beta}$ is easily 
calculable\cite{but1,but3} from Eqs. (\ref{dn}), (\ref{emissivity}), 
(\ref{micro}), and (\ref{ldos}),
\begin{equation}
D_{\alpha \beta} = \int d^3r \frac{(d\underline{n}_{\alpha}({\bf r})/dE)
 (d\bar{n}_{\beta}({\bf r})/dE)} {dn({\bf r})/dE} \ \ .
\label{d11}
\end{equation}

To illustrate the numerical procedure, we have computed the emittance of 
a T-junction quantum wire under the multi-mode and finite temperature
conditions.  The same system has been examined before for the single mode
and zero temperature situation\cite{wang1} using {\it externally} computed
GPDOS ({\it i.e.} Eq. (\ref{eq2})), thus the results here also
provide an useful comparison. As shown in Fig. (1), the wire has
two probes extending to $x=\pm \infty$ while the scattering region is 
provided by the T-junction as shown being bounded by the two dotted lines. 
We assume that the boundaries of this ballistic conductor are hard 
walls, {\it i.e.} the potential $V=\infty$. Inside the conductor the 
potential is zero. From now on we set $\hbar= 1$ and $m=1/2$ to fix our units.

Fig. (2) shows the emittance $E_{11}$ versus incoming energy $E$ in the 
first subband at zero temperature, where the GPDOS is computed using 
Eq.(\ref{gpdos}) (solid line) or using Eq.(\ref{eq2}) (dotted line).
The difference in the two curves comes solely from the difference in 
GPDOS. Notice that near the second subband the divergence in the dotted
curve is removed by using the locally computed GPDOS Eq.(\ref{gpdos}) as
shown by the solid curve. As discussed previously\cite{wang1} and shown in
Fig. (2), there is also a divergence near $E=E_1$ where $E_1=\pi^2$ is the
first subband energy, if using the {\it externally} computed GPDOS.  This
problem is also overcome using Eq. (\ref{gpdos}) as shown by the
solid curve. In fact since all quantities are computed locally if using
Eq. (\ref{gpdos}), there is no need for any correction terms to the DOS.
In Fig. (3a) we plot the emittance $E_{11}$ up to the 2nd subband energy 
for three different temperatures $T=0.005E_1$ (dotted line), $0.05E_1$
(dashed line), and $0.1E_1$ (solid line).  The inset shows the same 
quantity up to 3rd subband.  We have also depicted the dwell time 
$\tau_{d1}$ (dashed line) in Fig. (3b) together with the DC conductance 
(solid line) and the emittance $E_{11}$ at zero temperature (dotted line). 
There are eight resonant states located near the peaks 
of the dwell time. The behavior of the emittance $E_{11}$ near the 
resonant energies ${\cal E}_5$ (the 5th one) and ${\cal E}_6$ (the 6th)
which are basically inductive are different from that near the other 
resonant energies which are capacitive. This behavior is
useful in explaining the resonance behavior in the transmission
coefficient near resonant energies ${\cal E}_5$ and ${\cal E}_6$. For
instance, from the dwell time we know that near ${\cal E}_5$ there is a 
resonant state. However, it is not clear whether it is a resonant
transmission or resonant reflection since the DC conductance
has a sharp peak and a dip very close to each other. The fact that the 
emittance near ${\cal E}_5$ is inductive indicates that it is a resonant 
transmission. Finally, as the temperature is turned on, the
peaks or valleys in the emittance curve diminish gradually.  
When the temperature reaches $T=0.2E_1$ the interference pattern is washed 
out completely. This is expected since a finite temperature tends to smear 
out the quantum resonances.

In summary, we have proposed a numerical procedure and formula
for computing the global partial density of states which is precise for
any finite scattering volume of a quantum conductor. As GPDOS plays a most 
important role in the AC transport theory, our result provides a useful tool 
for further numerical investigations of the dynamic admittance.
In this formulation the electric current conservation is satisfied 
automatically. This formulation, especially Eq.(\ref{e6}), also applies 
to the non-linear transport\cite{but6,wang3}. 
Applying the new procedure to a T-shaped junction, the
divergences of the emittance at each subband edge are removed. 

\section*{Acknowledgments}

We gratefully acknowledge support by a RGC grant from the Government of 
Hong Kong under grant number HKU 261/95P, a research grant from the 
Croucher Foundation, the Natural Sciences and Engineering Research 
Council of Canada and le Fonds pour la Formation de Chercheurs 
et l'Aide \`a la Recherche de la Province du Qu\'ebec.
We thank the Computer Center of the University of Hong Kong for
computational facilities and the access of SP2 supercomputer.

\section*{Figure Captions}

\begin{itemize}

\item[{Figure 1.}] Schematic plot of the quantum wire system:
The wire width, the side-sub width, and height are fixed at $W=1$. The
two dotted lines separate the scattering region from the two probes. 

\item[{Figure 2.}] The emittance $E_{11}$ versus incoming energy $E$ in
the first subband at zero temperature, where the GPDOS is calculated
using Eq.(\ref{gpdos}) (solid line) or using Eq.(\ref{eq2}) (dotted
line). 

\item[{Figure 3.}] Figure (3a): the emittance $E_{11}$ versus energy for
three difference temperatures $T=0.005 E_1$ (dotted line), $0.05 E_1$
(dashed line), and $0.1 E_1$ (solid line), up
to the second subband. Inset, the same quantity up to the third subband.
Figure (3b): the conductance $G_{11}$ (solid line), the dwell time
(dashed line), and the emittance $E_{11}$ at zero temperature (dotted
line) where the dwell time and the emittance have been scaled by a
factor of eight to fit in
one figure. 

\end{itemize}

\begin{thebibliography}{00}
\bibitem{foot2}
For a pipe with a $\delta$-function potential $V({\bf r}) = \gamma
\delta({\bf r}-{\bf r}_0)$, the Green's function 
$G({\bf r},{\bf r'})$ can be derived analytically and is given by
$G({\bf r},{\bf r'}) = G_0({\bf r},{\bf r'}) +\gamma
G_0({\bf r},{\bf r}_0) G_0({\bf r}_0,{\bf r'})/(1-\gamma
G_0({\bf r}_0,{\bf r}_0))$ where $G_0$ is the Green's function
for a pipe with transverse wavefunction $\chi_m$. From
Ref.\cite{datta}, we have $G_0({\bf r},{\bf r'}) =\sum_m(-i/v_m)
\chi_m(y) \chi_m(y') exp(ik_m |x-x'|)$. In this case, one sees that
Eq.(\ref{relation}) is not satisfied because of the mode mixing.
\end{thebibliography}
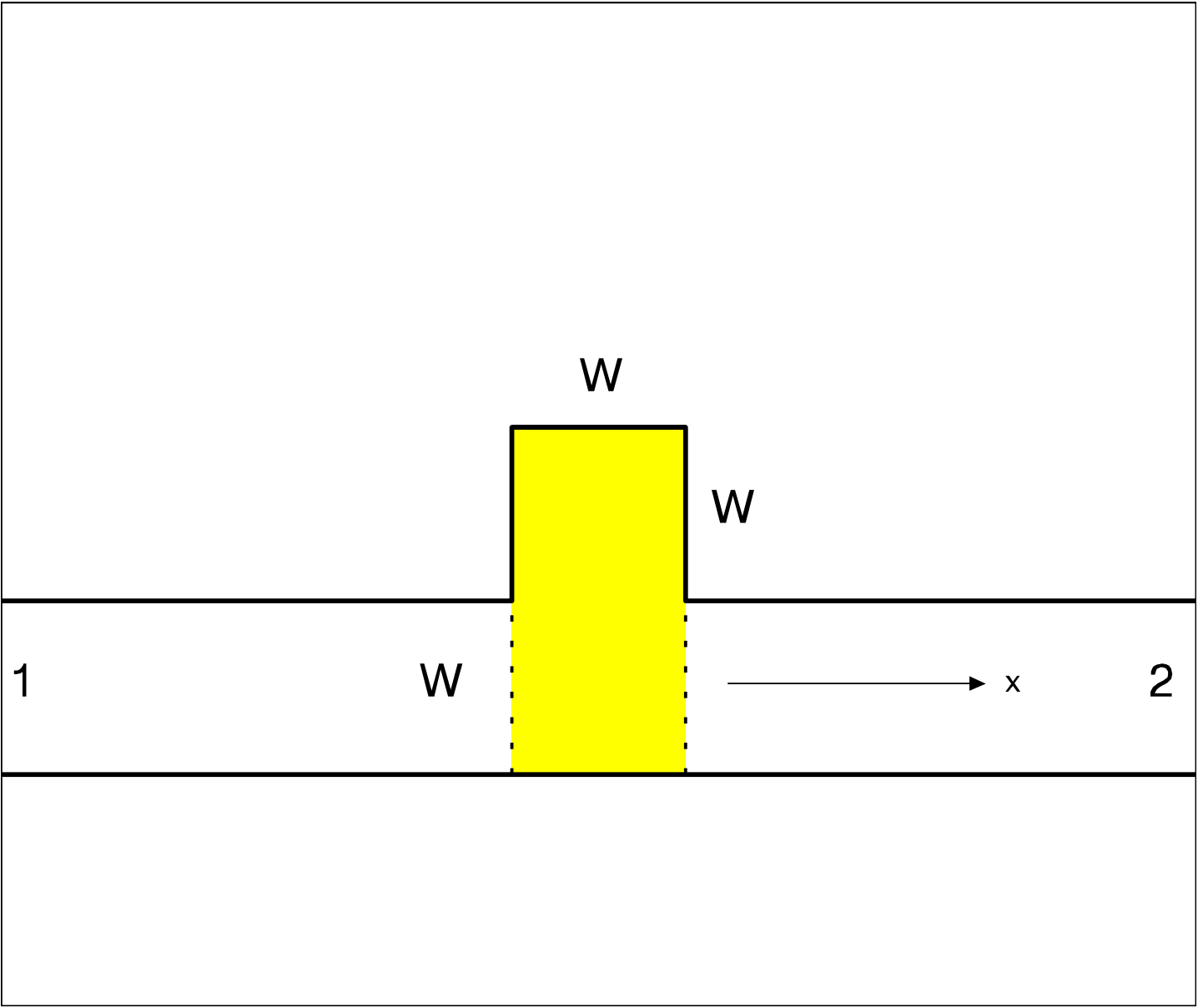
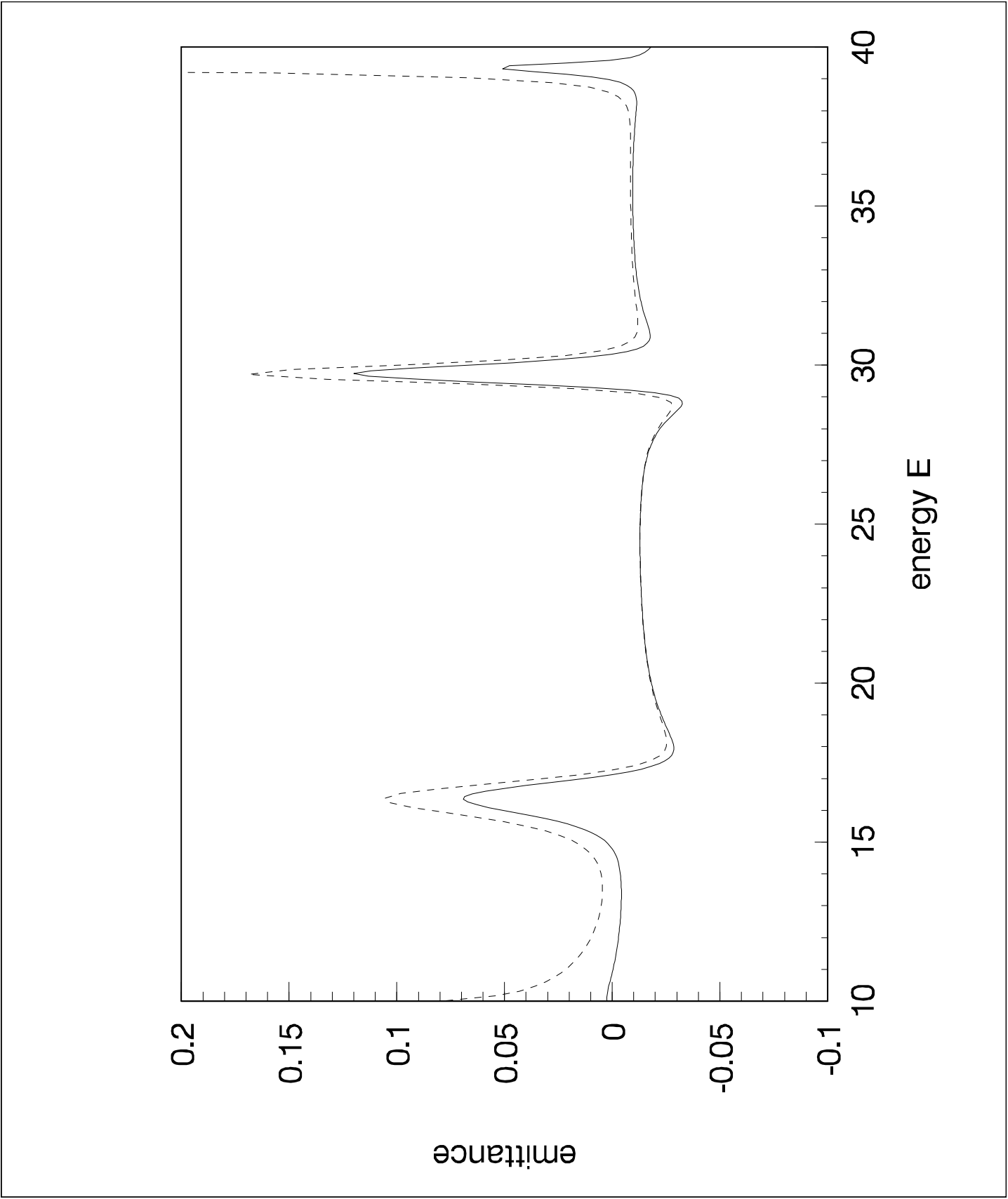
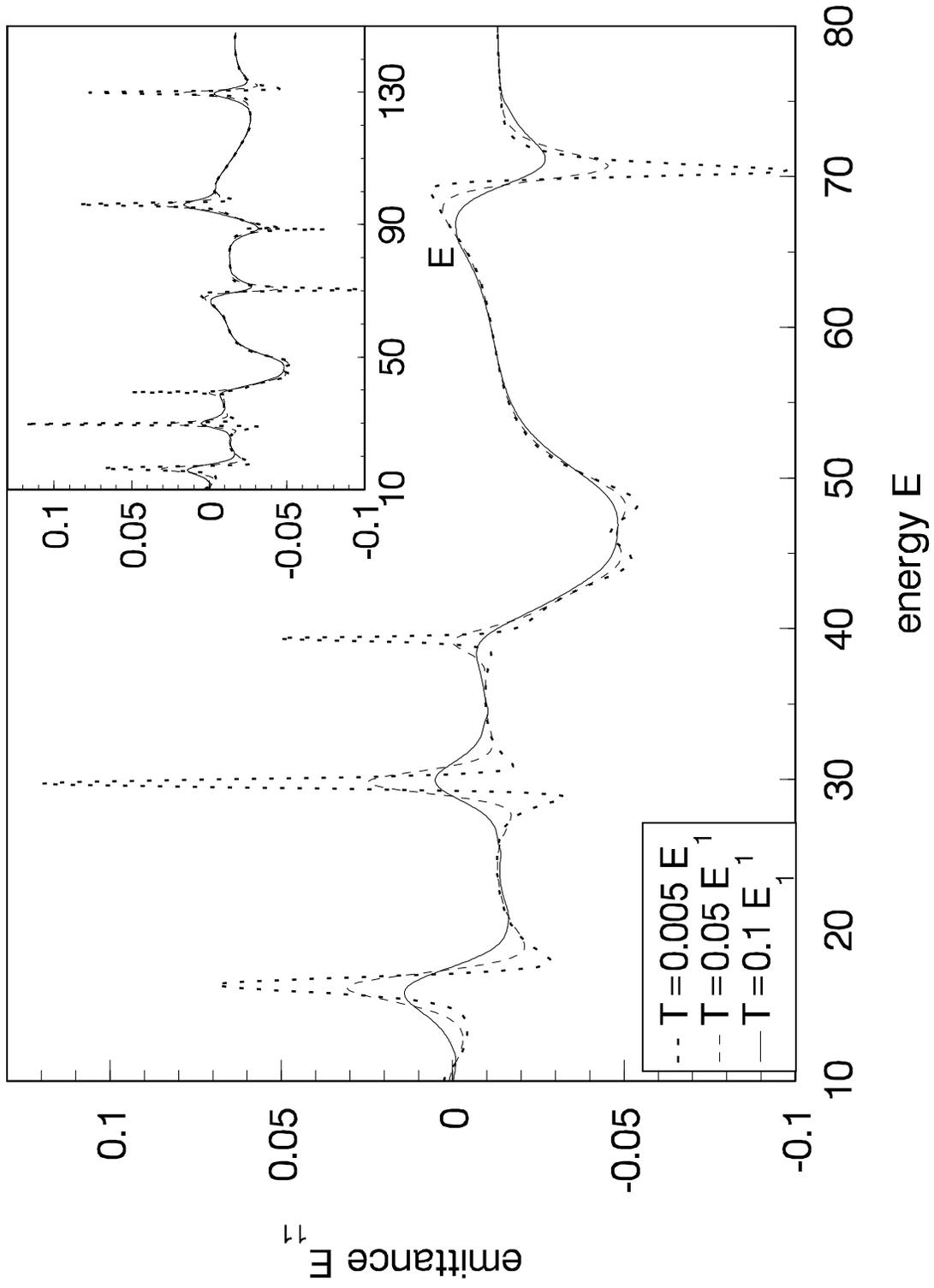
\end{document}